\documentclass[manuscript,natbib]{aastex}
\usepackage{natbib,subfigure}
\def\eso{ESO~328-IG~013}
\def\deg{\hbox{$^\circ$}}
\def\degb{\hbox{$^\circ$} }

\def\fs{\hbox{$.\!\!{}^{\rm s}$}}

\def\arcm{\hbox{$^{\prime}$}}

\def\s{$^{\prime\prime}$~}
\def\hh{$^{\mathrm h}$}
\def\mm{$^{\mathrm m}$}

\def\mjb{mJy beam$^{-1}$}

\def\arcs{\hbox{$^{\prime\prime}$}}

\shorttitle{Radio Polarization of SN 1006}
\shortauthors{Reynoso, Hughes, \& Moffett }

\begin{document}


\title{On the Radio Polarization Signature of Efficient and
  Inefficient Particle Acceleration in Supernova Remnant SN 1006}


\author{Estela M. Reynoso\altaffilmark{1,2}}
\affil{Instituto de Astronom\'\i a y F\'\i sica del Espacio (IAFE),
C. C. 67, Suc. 28, 1428 Buenos Aires, Argentina}
\email{ereynoso@iafe.uba.ar}

\author{John P. Hughes}
\affil{Department of Physics and Astronomy, Rutgers University, Piscataway,
NJ 08854-8019}
\email{jph@physics.rutgers.edu}

\and

\author{David A. Moffett}
\affil{Department of Physics, Furman University, Greenville, SC 29613}
\email{david.moffett@furman.edu}


\altaffiltext{1}{FCEyN, UBA, Argentina}
\altaffiltext{2}{Member of the Carrera del Investigador Cient\'\i fico of CONICET.}


\begin{abstract}

Radio polarization observations provide essential information on the
degree of order and orientation of magnetic fields, which themselves
play a key role in the particle acceleration processes that take place
in supernova remnants (SNRs).  Here we present a radio polarization
study of SN 1006, based on combined VLA and ATCA observations at 20 cm
that resulted in sensitive images with an angular resolution of 10
arcsec.
The fractional polarization in the two bright radio and X-ray lobes of the 
SNR is measured to be 0.17, while in the southeastern sector, where the radio 
and non-thermal X-ray emission are much weaker, the polarization fraction 
reaches a value of 0.6$\pm$0.2, close to the theoretical limit of 0.7.
We interpret this result as evidence of a disordered, turbulent
magnetic field in the lobes, where particle acceleration is believed
to be efficient, and a highly ordered field in the southeast, where
the acceleration efficiency has been shown to be very low.
Utilizing the frequency coverage of our observations, an average
rotation measure of $\sim 12$ rad m$^{-2}$ is determined from the
combined data set, which is then used to obtain the intrinsic
direction of the magnetic field vectors. 
While the orientation of magnetic field vectors across the SNR shell appear radial, 
a large fraction of the magnetic vectors lie parallel to the Galactic Plane. 
Along the highly polarized southeastern rim, the field is aligned tangent to the shock, and therefore 
also nearly parallel to the Galactic Plane. These results strongly suggest 
that the ambient field surrounding SN 1006 is aligned with this direction 
(i.e., from northeast to southwest) and that the bright lobes are due to a polar cap geometry.
Our study establishes that the most efficient particle acceleration
and generation of magnetic turbulence in SN 1006 is attained for
shocks in which the magnetic field direction and shock normal are
quasi-parallel, while inefficient acceleration and little to no
generation of magnetic turbulence obtains for the quasi-perpendicular
case.

\end{abstract}





\section{Introduction}

Shock waves driven by supernova remnants (SNRs) are proposed to
accelerate particles to relativistic energies through the diffusive
shock acceleration (DSA) mechanism
\citep{krymsky77,als77,bell78a,bell78b,blandford-ostriker78}.  In this
process, also known as first-order Fermi acceleration, electrons,
protons, and other ions scatter back and forth on magnetic
irregularities across the shock, gaining energy on each successive
crossing.  The orientation of the ambient magnetic field with respect
to the direction of the shock velocity and the level of turbulence of
the field throughout the shock transition region play key roles in the
DSA process \citep[for a review see][]{md2001}.  A critical issue of the DSA
theory concerns the amplification of the ambient magnetic field, which
is now believed to occur synergistically with particle acceleration
\citep{bl2001,bell}. In this picture, accelerated particles streaming
ahead of the shock excite magnetohydrodynamic turbulence that can
amplify the ambient magnetic field far above its initial seed value. The
field so produced is expected to be highly turbulent ($\delta B/B >
1$) \citep[e.g.,][]{bell}. Radio polarization observations are
therefore crucial to understanding such processes, since they can
provide information on the degree of order and orientation of the
magnetic field in the shocked gas of the remnant. Bilateral SNRs are
ideal targets for these studies---the highly symmetric morphology of
such remnants suggests evolution in a magnetized environment with
large-scale order.

The remnant of SN 1006, a large ($\sim 30$ arcmin) bilateral SNR
located at high Galactic latitude, is an ideal laboratory to study the
polarization and magnetic field distribution while avoiding confusion
from other sources in the line of sight.
It is widely accepted that SN 1006 is the remnant of a Type Ia
supernova event \citep[][p.~174]{stephenson-green02}.  Optical proper
motion measurements, combined with the shock velocity as inferred from
H$\alpha$ spectra \citep{gwrl02}, place this remnant at a distance of
2.18$\pm 0.08$ kpc \citep{wgl03}. At both radio and X-ray wavelengths
\citep[e.g.,][]{rg86, cassam+08,dyer09,roth+04}, the SNR appears as a
circular shell, with two bright lobes perpendicular to the Galactic
Plane. In contrast, the H$\alpha$ image \citep{wl97} reveals a system of
faint, relatively thin filaments defining an almost perfect, complete
circular ring, with enhanced emission at the NW, between the bright
radio and X-ray lobes.  Based on {\it XMM-Newton} X-ray observations,
\citet{abd07} estimated a higher ambient density at this location. TeV
energy $\gamma-$rays, originating in the radio and X-ray bright lobes,
have recently been detected with H.E.S.S.\ \citep{hess10}.

The first full polarization map of SN 1006 was obtained by
\citet{kundu} using 6 and 11-cm observations performed with the NRAO
43-m telescope at Green Bank.  The SNR was found to have an average
polarization percentage of 10\% to 15\% at 5 GHz and a radial
orientation of the magnetic field. A similar result was obtained by
\citet{dm76} using higher resolution observations, albeit with a lower
fractional polarization ($\sim 10$\%). The highest resolution
polarization study of SN 1006 was done by \citet{rg93}, based on VLA
observations at 1370 and 1665 MHz. They found peak polarization
fractions of 30\% in the lobes, somewhat higher than previous results,
and that the magnetic field, although presenting mostly a radial
orientation, was predominantly disordered.  Moreover, they also
detected weak polarized flux from the faint SE quadrant.

One of the current theories to explain the bilateral symmetry of SN
1006 \citep[e.g.,][]{fr90,dyer04,pet+09,schn+10} interprets the
limb-brightened emission of the lobes as a result of a shock expanding
into a region of the interstellar medium (ISM) whose magnetic field is
perpendicular to the shock normal. In the case of SN 1006, that means
that the magnetic field would be aligned SE to NW (the equatorial model). 
Acceleration rates are higher in perpendicular shocks
\citep{ebj95,jokipii}, as the magnetic field is directly compressed and
charged particles are accelerated as they traverse magnetic gradients;
however, the injection of thermal particles is lower due to the high
obliquity of fields with the shock normal, requiring a mechanism for
scattering thermal electrons and ions back into the shock. However,
other  work \citep[e.g.,][]{roth+04,cassam+08} on the variation
of the non-thermal synchrotron X-ray and radio emission with azimuthal
angle around the limb of SN 1006 casts doubt on this
interpretation. These results suggest that the morphology of the
emission is best explained if the bright limbs are polar caps,
requiring that the ambient magnetic field direction extends from SW to
NE. Acceleration rates in the bright caps in this situation
(quasi-parallel shocks) would be comparatively lower than in the
former case (quasi-perpendicular), but injection rates of thermal
particles would be much higher. The simple geometrical argument that
\citet{roth+04} applied to the {\it XMM-Newton} image, favors the
equatorial belt model when applied to a radio image. To reconcile the
model with the observations in both spectral ranges, \citet{petruk+2011}
propose a somewhat contrived model of a magnetic field with a
gradient.

When DSA is very efficient, the back reaction of the accelerated
particles on the structure of the supernova remnant can have
significant and observable consequences. One of these, potentially
active in young ejecta-dominated remnants, is a narrowing of the
distance from the forward shock to the contact discontinuity
\citep{decouchelle00}, an effect that was first observed in the Tycho
SNR \citep{warrenetal05}. \citet{cassam+08}, using {\it Chandra} X-ray
observations, discovered this effect in SN 1006 by comparing the
outermost location of the H$\alpha$ filaments to the outermost extent
of the ejecta, as traced by the 0.5-0.8 keV band which is dominated by
emission from highly ionized O ions in the SN ejecta. This study found
that the gap between the forward shock and contact discontinuity was
largest at the SE rim and that the gap smoothly decreased with
azimuthal angle as one proceeded toward the bright lobes.  The authors
concluded that this was direct evidence for a variation in
acceleration efficiency as a function of azimuth around the rim of SN
1006, likely due to a corresponding azimuthal variation in the
orientation between the shock normal and the ambient magnetic field
direction.

In this paper we offer insights on these critical issues by carrying out 
a new high resolution polarization study of SN 1006 based on observations 
performed with the Australia Telescope Compact Array (ATCA) and the Very Large 
Array (VLA). The combination of these two data sets allows us to construct 
images with unprecedented detail, from which new information is derived to 
clarify the role of magnetic fields in the emission of SN 1006.

\section{Observations and Data Reduction}

Full-polarization radio continuum observations of SN 1006 were carried
out in 2003 using the Very Large Array (VLA) of the National Radio
Astronomy Observatory and the Australia Telescope Compact Array
(ATCA).  VLA observations were performed on 24 January 2003 for a
four-hour duration in its CnD array configuration (primarily used to
observe sources at southern
declinations). Visibility data were recorded from two 12-MHz frequency
bands centered at 1370 and 1665 MHz. We used the source 3C 286 as our
flux calibrator, and 1451$-$400 as the phase calibrator for SN 1006.
ATCA observations were performed on three separate occasions: 12 hours
in the 6B configuration on 24 January, 12 hours in the 6A
configuration on 3 March, and seven hours in the 750C configuration on
12 June.  Visibility data were recorded from two 128-MHz frequency
bands centered at 1384 and 1704 MHz, divided into 32 channels each.
We used PKS 1934$-$638 as our flux calibrator, and PKS 1458$-$391 as
the phase calibrator.

Flux, phase and polarization calibration of VLA data was performed
using the Astronomical Image Processing System (AIPS). The ATCA data
was calibrated using the Miriad software package \citep{sault+95}.  We
used Miriad to perform the remainder of the reduction process. The
Stokes visibility data from all observations were split into two data
sets: 1370 MHz VLA data were combined with 1384 MHz ATCA data in one
set, and 1665 MHz VLA data were combined with 1704 MHz ATCA data to
form the other set.  We created uniform-weighted `dirty' images of
Stokes I (total intensity), Q and U for both frequency data sets using
the Miriad task INVERT.  Images were mosaicked to handle a slight
pointing offset between the VLA and ATCA observations. The
Miriad task PMOSMEM was used to perform a joint maximum-entropy deconvolution
of the mosaicked dirty images.  This program can operate on all Stokes
images handling the positive and negative emission found in Stokes Q
and U maps.  Although the resulting resolution of the dirty maps was
$\sim$ 8\s $\times$ 6\s, we restored them during deconvolution with a
circular beam of 10\s to improve their signal-to-noise.  The total
intensity image constructed with these data was previously published
in \citet{cassam+08}.

We used the Miriad task IMPOL to combine the Stokes $I$, $Q$ and $U$ to form 
images of linear polarized power ($I_P=\sqrt{Q^2+U^2}$), polarized fraction 
($I_P/I$), linear position angle ($\psi = 1/2$ tan$^{-1} (U/Q)$), a rotation 
measure map between frequency data sets, and all of their associated error 
maps. All polarization maps (except total intensity) were blanked where $I_P$ 
was lower than $2\sigma$, where $\sigma$ is the noise level for the Stokes $Q$
and $U$ maps. We used $\sigma_Q = \sigma_U = 0.085$ \mjb. The Ricean bias was 
removed in first order in the same task by computing the polarized power as 
$\sqrt {Q^2 + U^2 - \sigma^2}$, hence the minimum value of the polarized 
emission is $\sim$ 0.15 \mjb. In Fig. 1 we present the resulting total 
intensity (1(a)) and polarized intensity (1(b)) maps of SN 1006, where in the 
latter, pixels where the total intensity was lower than 0.25 \mjb \ (the 
outermost contour of SN 1006 according to Fig. 1(a)) were further blanked.

\begin{figure}[ht]
  \centering
  \subfigure[Total intensity]{
    \label{total}
    \includegraphics[width=7.5 cm]{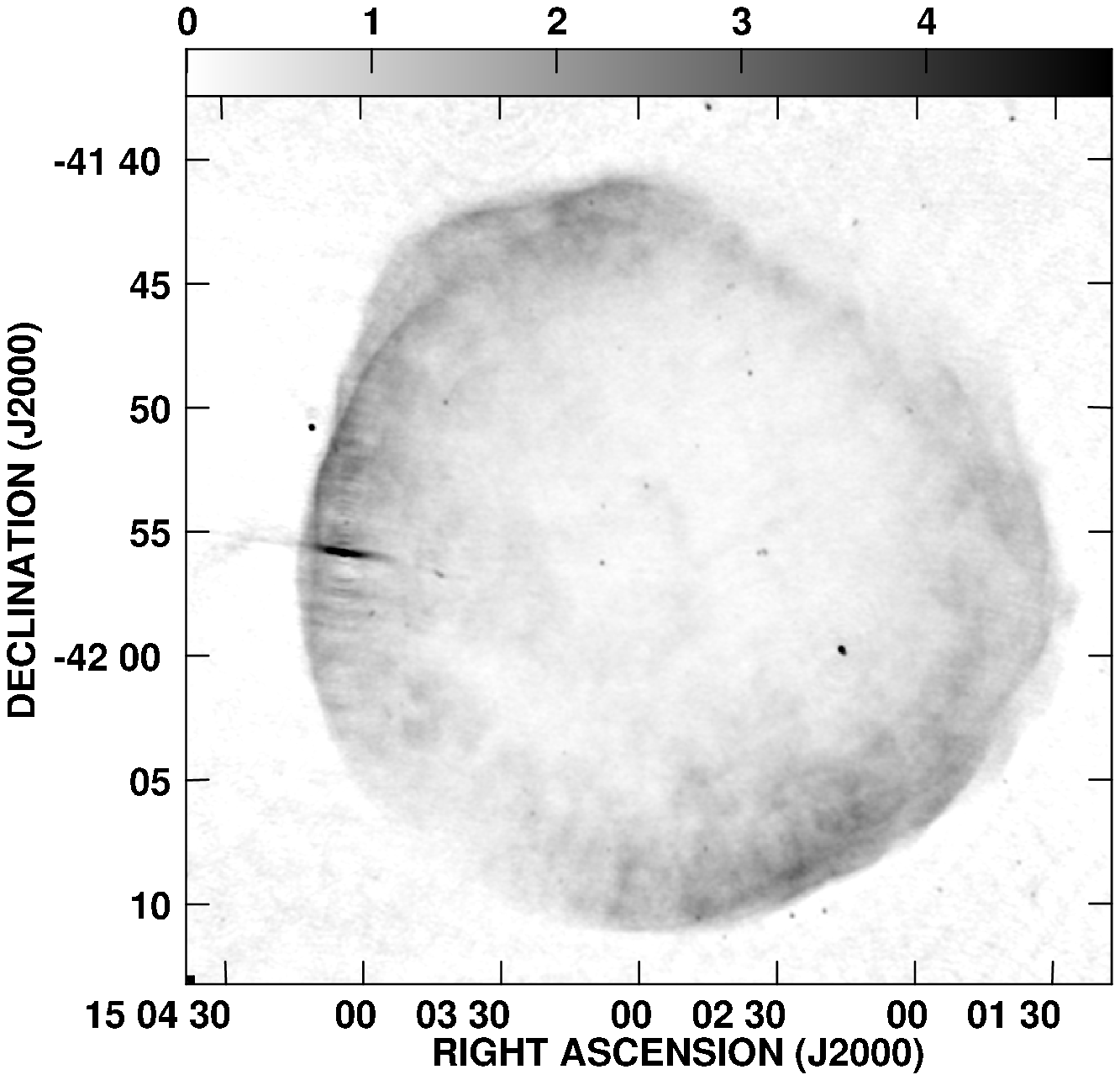}
  }
  \subfigure[Polarized intensity]{
    \label{pol}
    \includegraphics[width=7.5 cm]{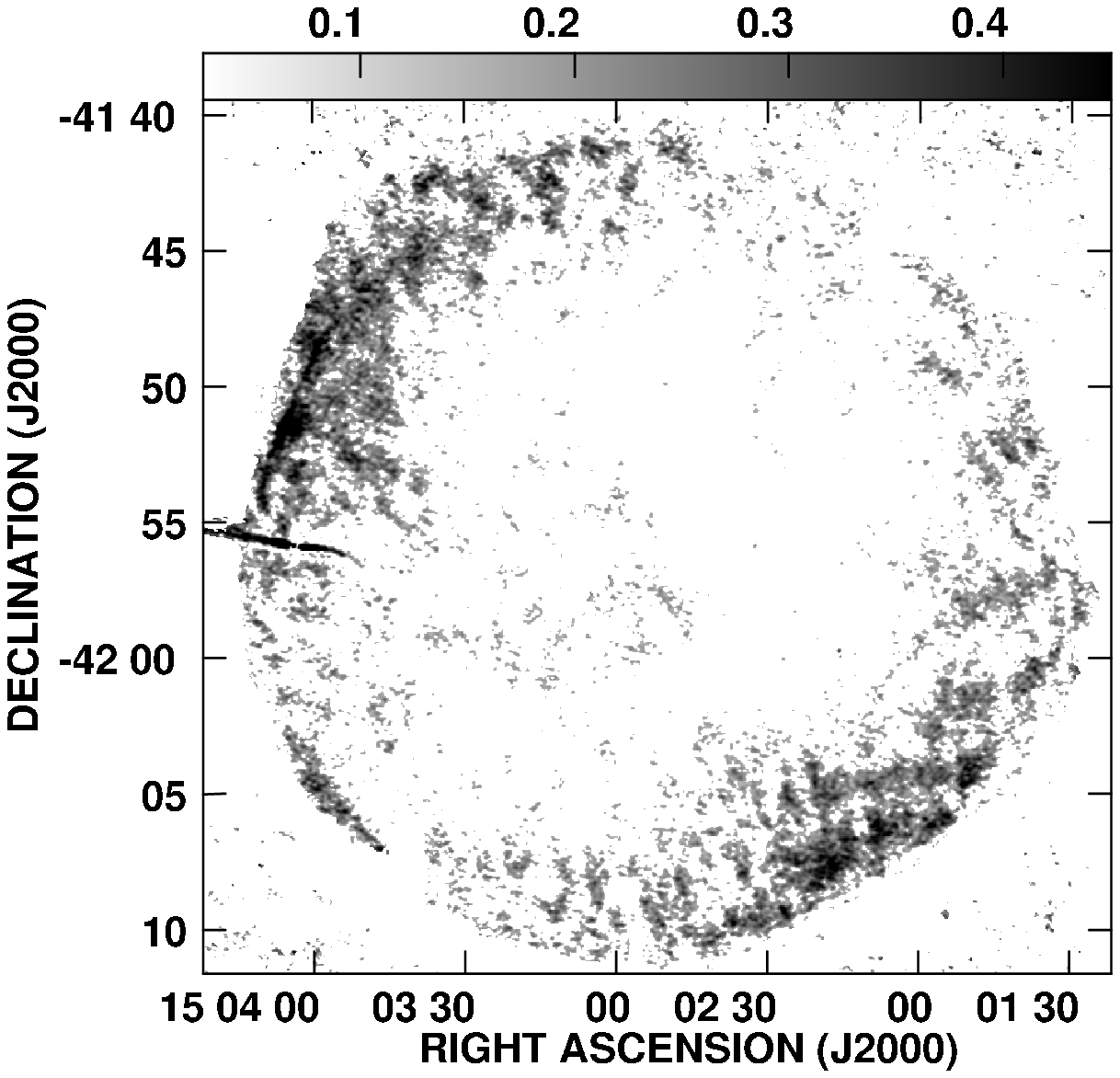}
  }
  \caption{(a) Total intensity image of SN 1006 at 1.4 GHz. The intensity
scale, in \mjb, is displayed on top of the figure. The beam size is 10 arcsec.
(b) Polarized intensity at 1.4 GHz. The intensity greyscale and resolution are
as in (a), although the former with a different scale.} 
  \label{intens}
\end{figure}

The integrated flux density of our total intensity map is $\sim 17$
Jy, in agreement with \citet{cassam+08}, where an independent
calibration of the same data was presented. This value is lower than
the combined single-dish and interferometer flux of 18.68 Jy found by
\citet{dyer09} at the same frequency.  We expect a shortage
due to missing flux from short {\it uv} spacings.  We also recovered
an integrated flux of $\sim 2$ Jy from our linear polarization map
after primary beam correction, in excellent agreement with the VLA
observations of \citet{rg93}. Their images also include central
emission in excess of that found in our images.  Indeed, the
integration time, bandwidth and {\it uv} coverage of differing
observations, array configurations, and interferometers can result in
images with different background noise levels and integrated flux
densities, but we must also note here that the use of different image
processing techniques can too. To verify the integrity of our process
and resulting images, we processed our VLA and ATCA data separately,
and found that the polarization maps were also similar. We are
confident the combined reduction did not corrupt the resulting
polarization maps, and the combination of VLA and ATCA visibility data
actually improved the overall sensitivity toward small-scale
structure, as can be appreciated, for example, towards the background
source \eso\ (section 3.1).

\begin{figure}[ht]
  \centering
    \includegraphics[width=6.8 cm]{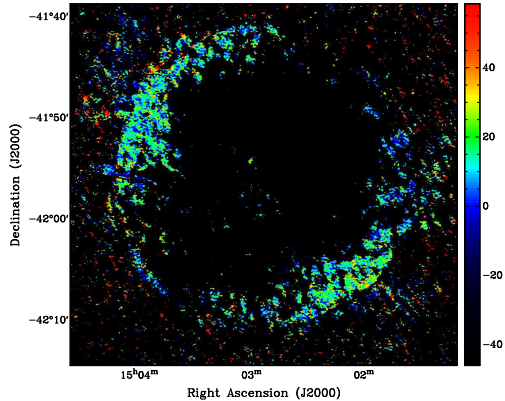}
  \caption{Distribution of the rotation measure on SN 1006, at 10
arcsecs resolution. The scale bar, shown at the right, is given in radians
per square meters.}
    \label{rotmeas}
\end{figure}

As mentioned above, the polarization position angle maps at the two
frequencies were combined by IMPOL to form a rotation measure (RM) map
that could be used to derotate the position angles in order to derive the
intrinsic direction of the electric field vectors.  Earlier work by
\citet{dm76} and \citet{rg93} show that the RM toward SN 1006 does not
cause position angles to wrap more than 2$\pi$ radians, so we computed
the RM for each pixel using only two frequencies, assuming no $\pi$
radian ambiguities. The resulting image is shown in Fig. \ref{rotmeas}.  We 
found that the distribution of RM values across both the NE and SW lobes of the 
shell is well represented by a Gaussian distribution centered at 12 
rad m$^{-2}$, with a $1\-\sigma$ deviation of 20 rad m$^{-2}$.  This level of 
deviation is expected, considering that the use of a $2\-\sigma$ cutoff in the 
Stokes Q and U maps corresponds to a position angle error of 14 degrees.  
Inspecting each lobe separately, we find that the RM distributions peak at 12 
rad m$^{-2}$ and 14.5 rad m$^{-2}$ for the NE and SW lobes respectively. We 
found no evidence of significant differences 
of the RM value in each lobe, unlike that found in G296.5+10.0 \citep{h-s+10}, 
another high-latitude bi-polar SNR whose RM varies significantly, from $-92$ in
one of the lobes to +44 rad m$^{-2}$ in the other one. In G296.5+10.0, the 
difference is attributed to a toroidal magnetized wind from the progenitor, 
which is not the case for SN 1006, where the similarity in the RM values for 
both lobes is quite reasonable considering the symmetry of the source.

\begin{figure}[ht]
  \centering
    \includegraphics{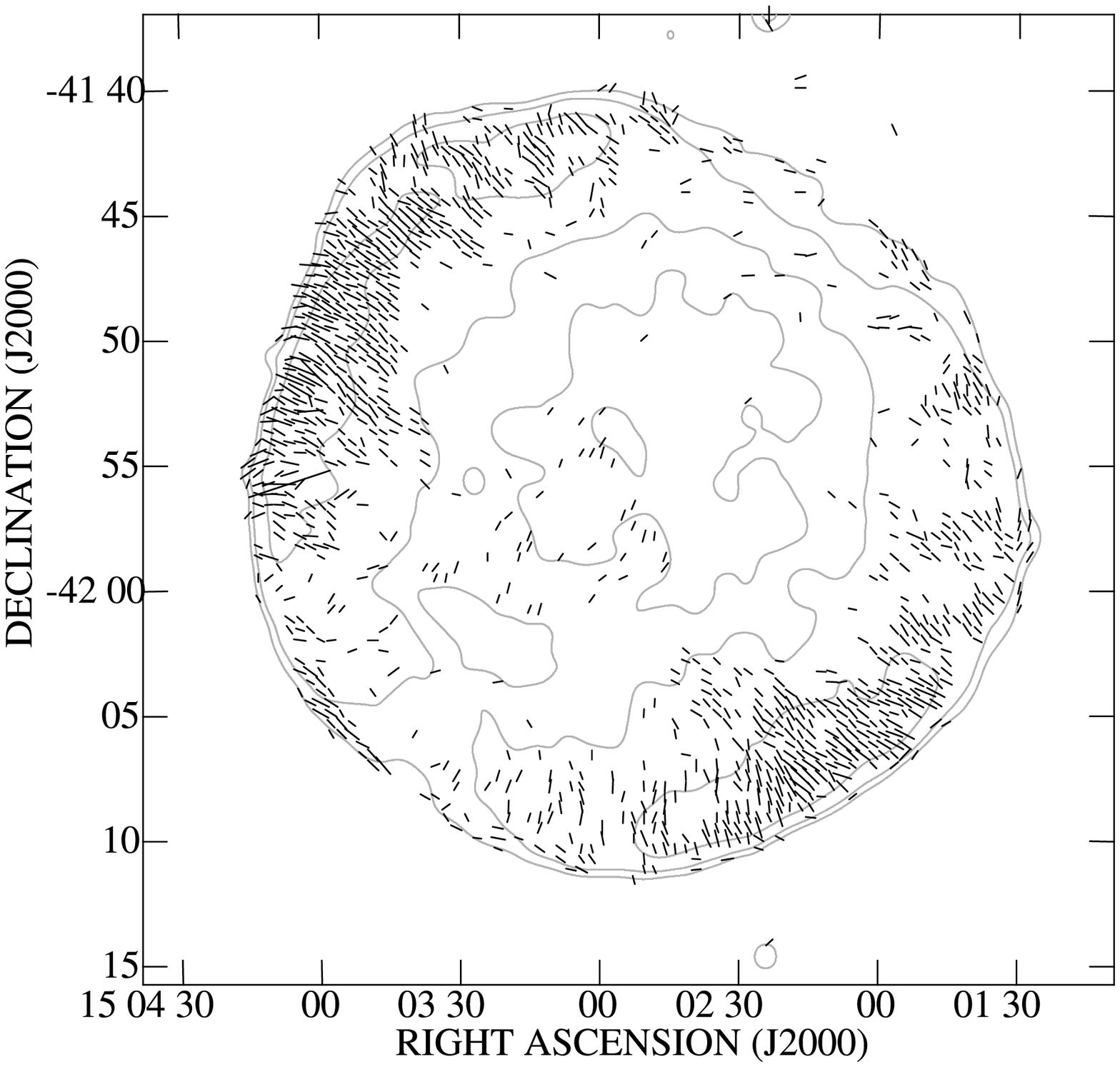}
  \caption{Distribution of magnetic field vectors on SN 1006 at 1.4 GHz 
corrected for Faraday rotation (assuming uniform RM=12 rad m$^{-2}$), at 10 
arcsecs resolution. Total intensity contours at 10, 20 and 50 \mjb, are 
superposed, where the beam was convolved to 60 arcsecs. For the vectors, a 
length of 30 arcsecs represents 0.25 \mjb \ of polarized flux.} 
    \label{magvec}
\end{figure}

Our RMs agree with previous polarization studies within the quoted errors.  
\citet{rg93} found an RM of $28 \pm 16$ rad m$^{-2}$ on the NE lobe, and 
$16 \pm 14$ rad m$^{-2}$ on the SW lobe.
A single-dish polarization study by \citet{m71} at 2.7 and 5.0 GHz yielded RM 
values of $21 \pm 5$ rad m$^{-2}$ in the NE, and $9 \pm 5$ rad m$^{-2}$ in the
SW, although a re-analysis of similar data by \citet{dm76} had an error of 
23 rad m$^{-2}$.  While the RM appears different on average between the two 
lobes, they are not significantly so.  Using the assumption that our average 
RM value of 12 rad m$^{-2}$ was uniformly distributed over the remnant, the 
lowest frequency polarization position angle map was derotated to the position 
angle at zero frequency, and further rotated 90\deg \ to display the
orientation of the magnetic field. The resulting magnetic vector map
is shown in Fig.~\ref{magvec}. Pixels were blanked where the flux
density fell below 10 \mjb \ in a total intensity image convolved to
60 arcsec. The magnetic vectors show an overall radial direction, and
appear very much the same as reported by \citet{rg93}.

\medskip

\section{Results}

\subsection{Fractional polarization}

\begin{figure*}[ht]
  \centering
  \includegraphics{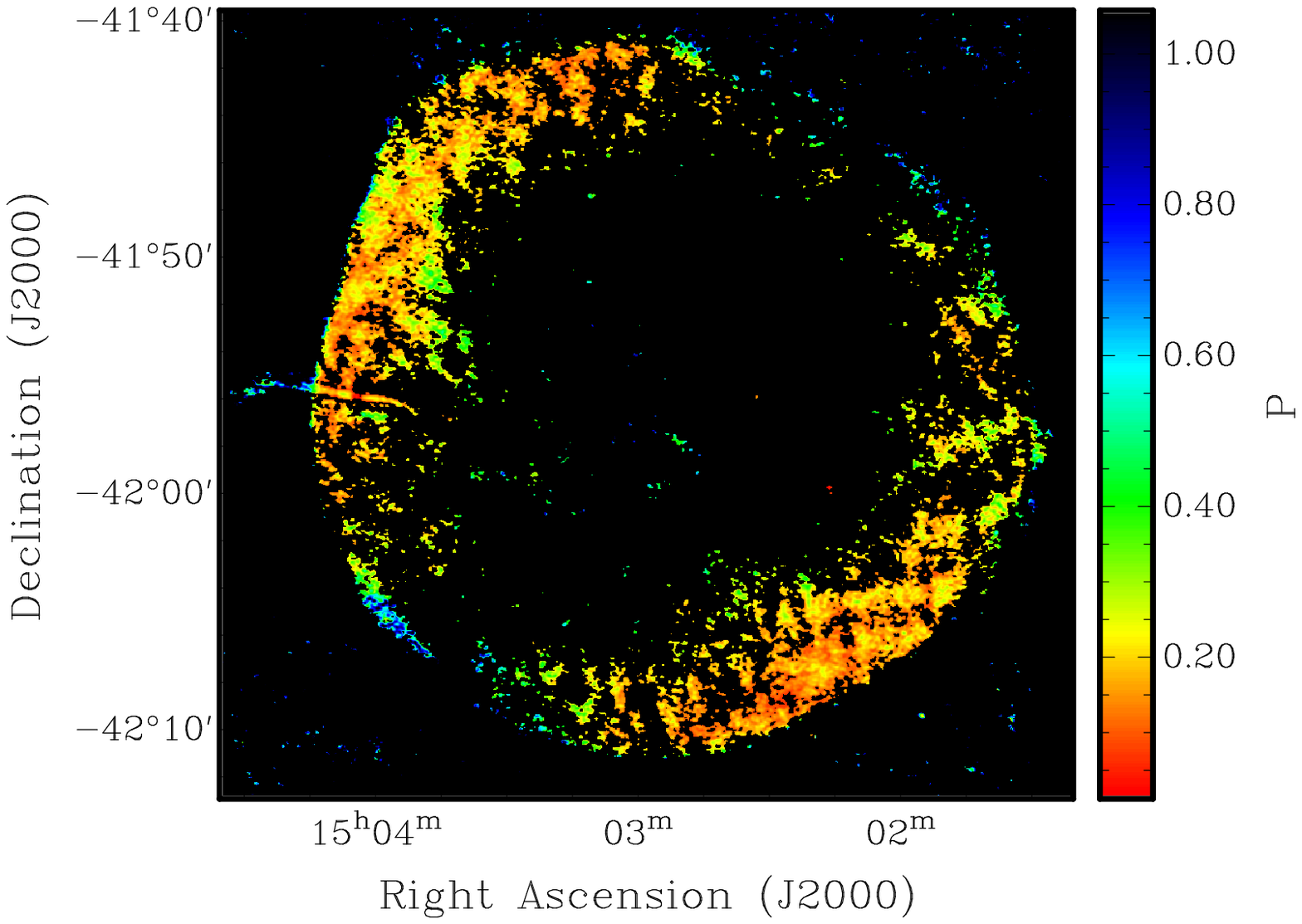}
  \caption{Fractional polarization $p$ of SN 1006 at 1.4 GHz. The resolution is
10 arcsecs. The color scale is shown at the right. Only pixels where $p$ was
at least twice its error were kept.} 
  \label{fracpol}
\end{figure*}

Figure \ref{fracpol} shows the distribution of the fractional polarization $p$,
computed as described in the previous Section. The $p$ map was compared to the
error map $\sigma_p$, and those pixels for which the ratio $p/\sigma_p \leq 2$ 
were blanked. In this image, unpolarized emission is represented in red. 
The average fractional polarization in the SE and NW lobes is $17 \pm 7\%$ . 
\citet{rg93} note that since their images do not 
include single dish observations, a smooth component of the total intensity is 
missing, so that if all the polarized flux is assumed to be detected, the 
fractional polarization measured is only an upper limit. They discuss this 
problem and conclude that the ratio maps must show values very close to the 
true ones, especially near the rim. Since the total flux we measure is much 
closer to the real one, the fractional polarization should be accordingly 
closer to its true value. 

\begin{figure}[ht]
  \centering
 \resizebox{\hsize}{!}{\includegraphics{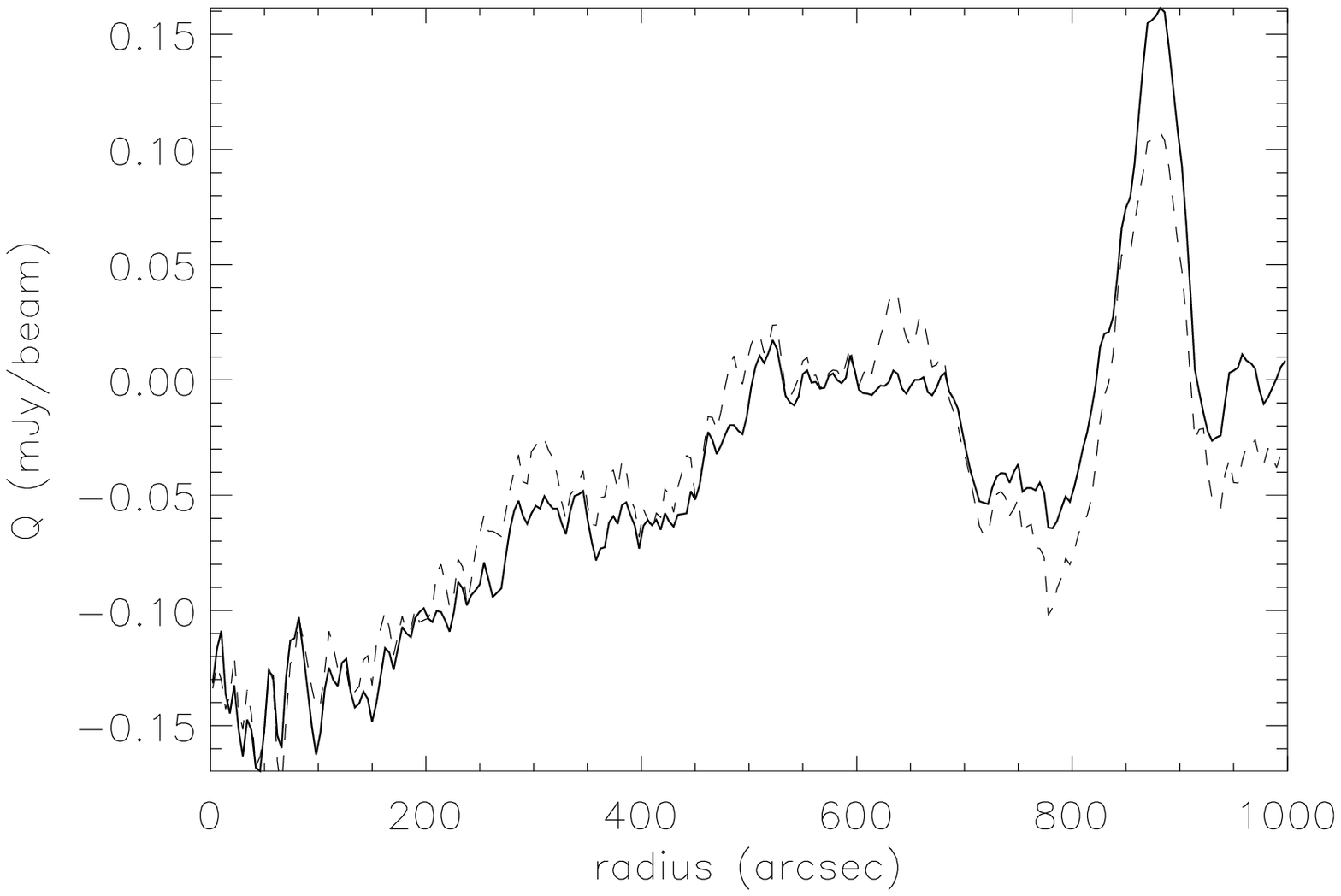}}
  \caption{Azimuthal average of the Stokes parameter Q computed between
113\degb and 180\degb over annuli of 4 arcsec width. The dashed curve
shows the same but averaged between 140\degb and 180\deg, where the
feature at 880 arcsec is weaker.} 
  \label{qplot}
\end{figure}

\citet{rg93} called attention to a faint arc on the SE rim visible in
the Stoke parameter Q map that closely follows the circular rim of the
shell. The authors argued that this feature is real, however their
sensitivity did not allow them to provide additional information about
its morphology or polarization properties. We have examined our Q
image and found that the highest emission occurs at an arc that
extends from 113\degb to 180\degb at a radius of 880 arcsec centered
at RA(2000)= 15\hh 2\mm 53\fs36, Dec.(2000)=$-41^\circ$ 56\arcm
25\arcs. Between 113\degb and 140\deg, the emission at the arc is
several times above the noise, while the rest of the feature is
weaker. In Fig.~\ref{qplot} we show the average emission in the Q
parameter on 4-arcsec wide annuli from the center over the sector
defined by the full arc. The azimuthal average clearly enhances the
emission at the arc, confirming that this feature is real. The same
result is observed when only the weakest part of the arc is considered
(dashed line in Fig.~\ref{qplot}). In Fig.~\ref{fracpol}, this feature
appears to be highly polarized, with $p=0.6 \pm 0.2$, where the
uncertainty is the RMS of the observed polarization fractions rather
than the error on the mean. High polarization is also observed at the NW 
rim. These results will be discussed in the next section.

\begin{figure}[ht]
  \centering
 \resizebox{\hsize}{!}{\includegraphics{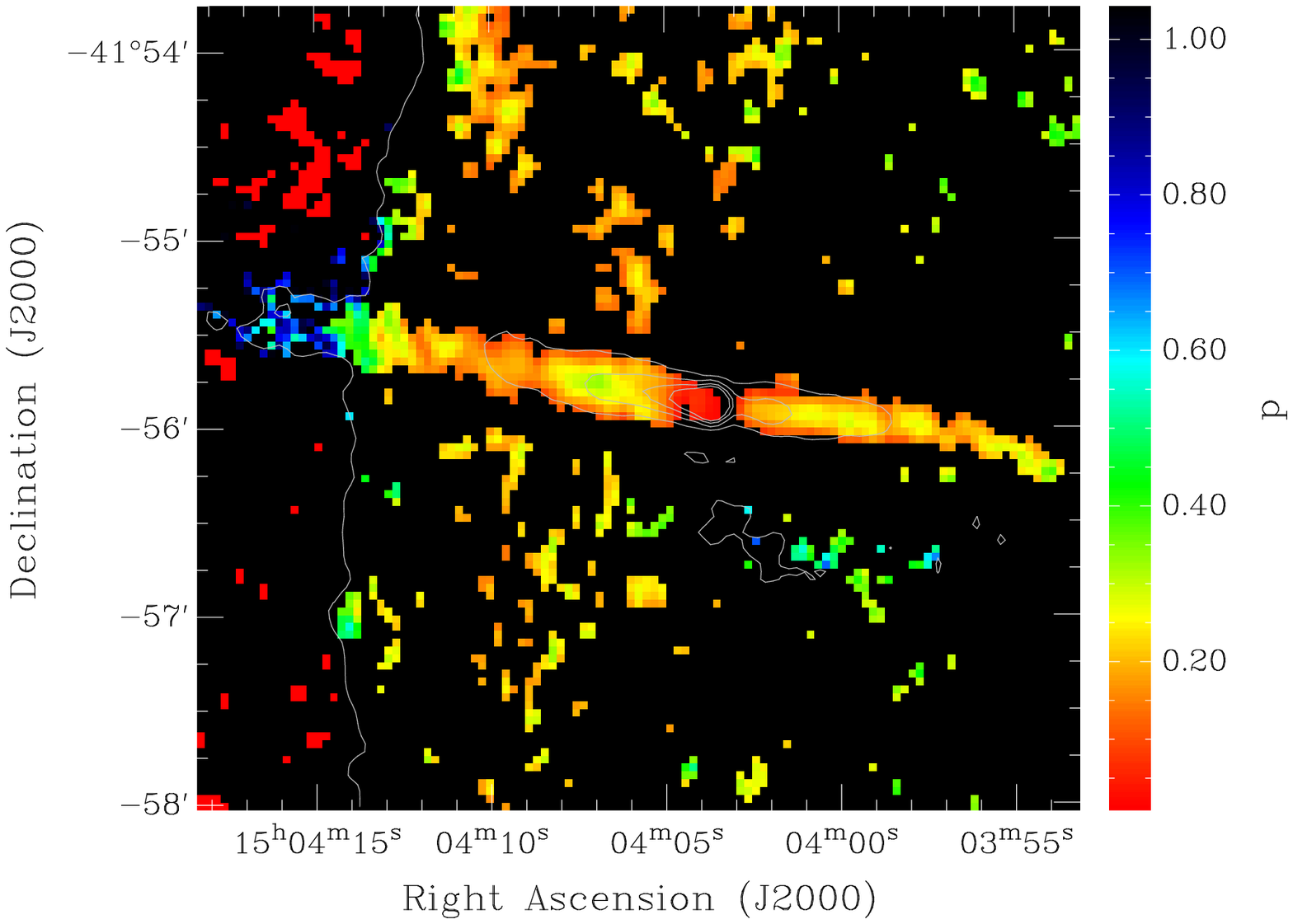}}
  \caption{Fractional polarization of the extragalactic source \eso,
located on the NE lobe of SN 1006. The resolution is $\sim 8\arcs \times 
6\arcs$. A few radio continuum contours are included to show the outer rim of 
the SNR.} 
\label{galaxy}
\end{figure}

Finally, these observations allowed us to measure the polarization of
a bright extragalactic source that lies on the remnant's shell at a
position angle of about 90\deg. This source is cataloged in NED as
\eso, an interacting galaxy \citep{Lau82}; it is also coincident with
a 2MASS galaxy (2MASX J15040364$-$4155511) at a redshift of $z=0.039$
and is also likely the radio source PMN J1503$-$4145. \citet{rg93}
noted that the fractional polarization was too high to be
interpreted as the core of a radio galaxy or quasar. In fact, they
found a fractional polarization of about 11\% because they could not
resolve the core of this elongated source.  Our data reveal two
polarized jets emerging from a clearly unpolarized core
(Fig.~\ref{galaxy}).  The jets extend for at least 2$^\prime$ ($\sim$100 kpc)
%
%
and have typical polarizations of 18\%$\pm$5\% on the shell of SN 1006
and 65\%$\pm$15\% beyond the rim, and the core has $p=$3\%$\pm$0.8\%,
more in line with the value expected for the core of an extragalactic
source \citep[e.g.,][]{saikia}. The high polarization of that portion
of the jet that lies beyond the rim is probably close to its true
intrinsic value; however, over most of its extent the jet is superposed 
on the SNR emission which makes the fractional polarization appear lower 
than its true value. To support this argument, we note the remarkable 
difference between the flux density of the galaxy just to the East of the 
edge of SN 1006 ($\sim 0.65$ \mjb ) and immediately to the West ($\sim 2$ 
\mjb), while the polarized flux for the same regions is about 0.46 and 0.52 
\mjb \ respectively. Moreover, the flux of SN 1006 just to the North and
South of the galaxy in the inner border of the shell varies between 1.3 and
1.5 \mjb, implying that there is a major contribution from the SNR to the 
flux measured on the galaxy. The contribution from the SNR decreases slowly
along $\sim$ 1 arcmin inwards from the SNR edge, and much faster as it gets
closer to the core of the galaxy. At the core, the flux density of the
galaxy is 20 times higher than the SNR. A detailed study of the polarization 
measurements on \eso\ will be the subject of a separate paper.

\subsection{Magnetic Field Orientation}

Figure \ref{magvec} shows the distribution of magnetic vectors after
combining the ATCA and VLA polarization observations. A comparison
with Fig.~8 in \citet{rg93}, where only VLA observations were used,
shows a close agreement in those regions where data are available,
although they detected higher levels of polarized emission in the
center of the remnant. Based on previous studies, in which polarized
emission could only be detected on the brightest lobes, it was
suggested that the magnetic fields were predominantly radial. There
were, nevertheless, hints from \citet{rg93} of a non-radial field
orientation along the faint SE rim based on two, barely significant,
magnetic field vectors. Our data allow us to gain insight on the
details of the magnetic field distribution at the outer rim and at the
lowest intensity regions in the SE and NW.

\begin{figure}[ht]
  \centering
  \subfigure[Magnetic vectors]{
    \label{polvec}
    \includegraphics[width=10 cm]{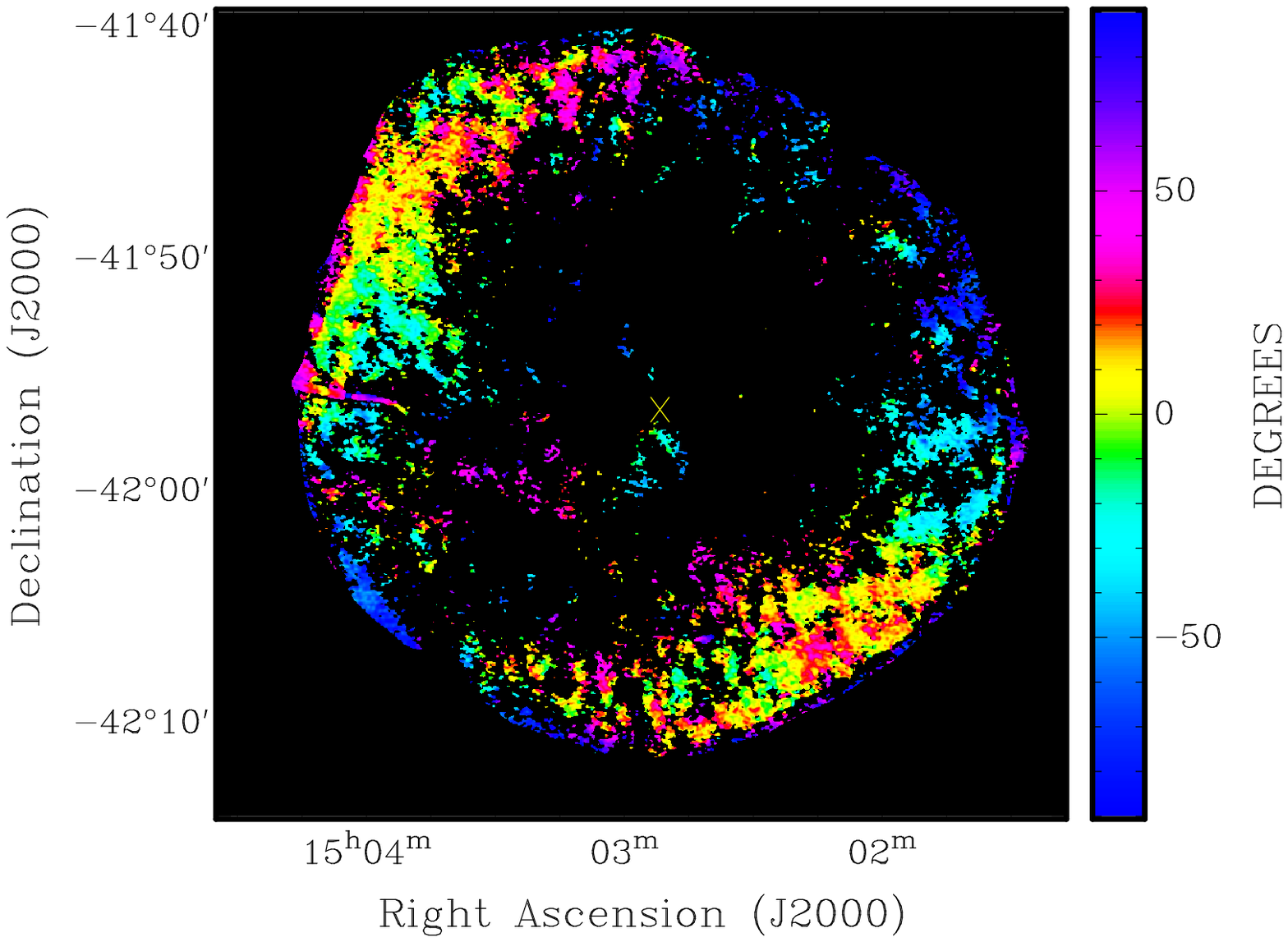}
  }
  \subfigure[Radial and fixed angle distributions]{
    \label{overlap}
    \includegraphics[width=9 cm]{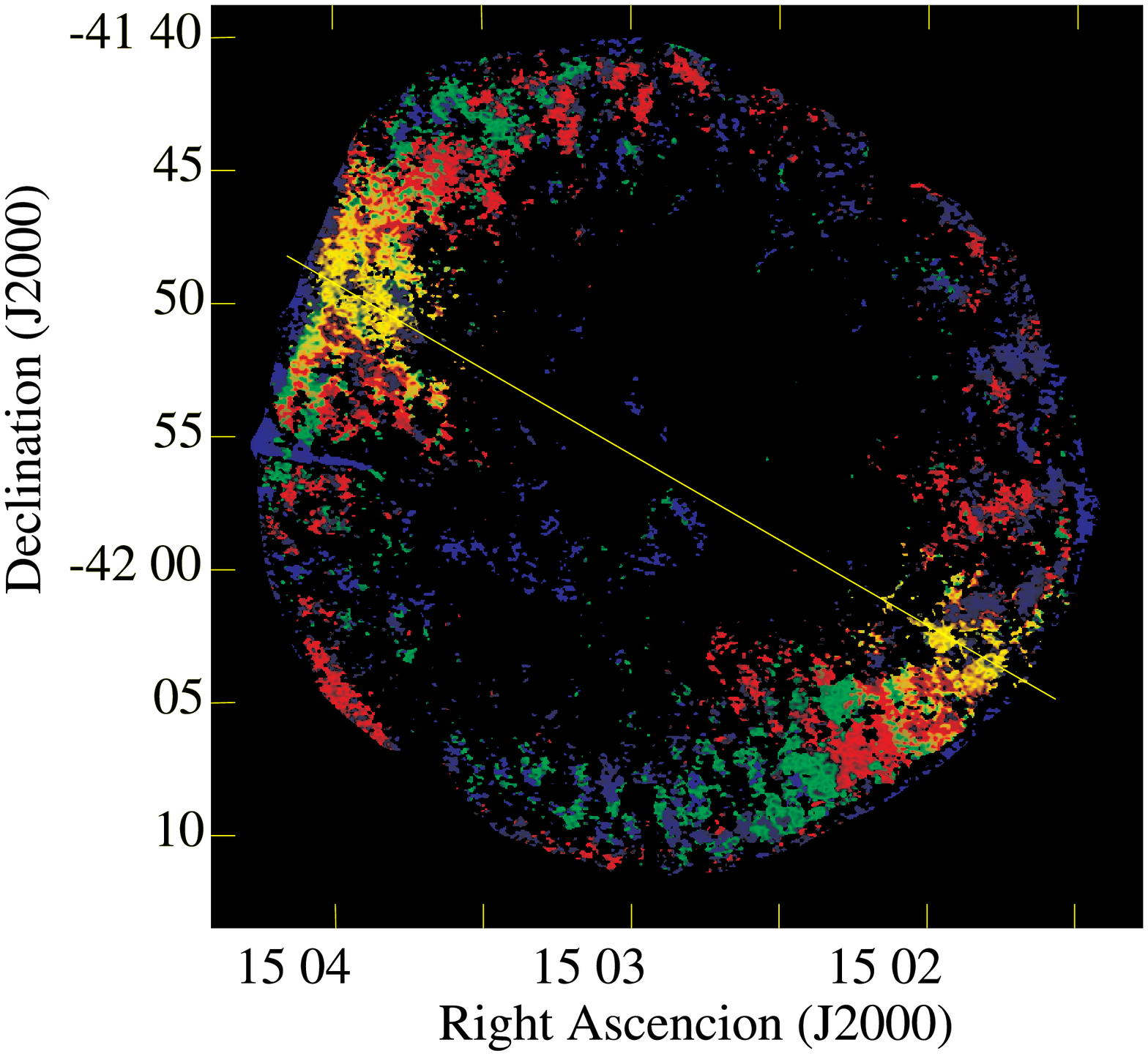}
  }
  \caption{(a) Magnetic field orientation with respect to polar angle (polar-referenced angle).  The center of the polar coordinate system used to define the polar angle (local radial direction) is marked by a yellow cross at the center of SN 1006. The color scheme of the legend is cyclic; blue represents both $90^\circ$ and
$-90^\circ$. A positive polar-referenced angle indicates a counter-clockwise angular difference between magnetic vectors displayed in Fig.~\ref{magvec} and the polar angle.
(b) Magnetic field orientation with respect to the Galactic Plane and polar angle.  Red pixels are for vectors at a fixed angle of $60^\circ$ (the direction of the Galactic Plane), while green indicates  
vectors that are locally radial. In both cases, a tolerance of $\pm 14^\circ$ 
is adopted with the intensity of pixels fading as they approach the limit.
Pixels that do not fall in any of these two groups are plotted in blue, such 
that the fainter the blue, the closer to $60^\circ$. The yellow line indicates the 
direction of the Galactic Plane.} 
  \label{rad_gp}
\end{figure}

In order to determine whether the orientation of the magnetic field is
radial along the whole SNR, we computed the angular difference between 
each magnetic vector and the radial direction, a polar-referenced angle, by adopting a polar coordinate 
system centered at the SNR's nominal center, located by 
\citet{rg86} and \citet{wl97} at RA(2000)= 15\hh 2\mm 51\fs7, Dec.(2000)=$-41^\circ$ 56\arcm
33\arcs. The polar-referenced angle varies between 90$^\circ$ and $-90^\circ$;
it is negative for a clockwise shift with respect to the
radial direction and positive for a counter-clockwise shift.  
Since the two extreme angular values represent the
same orientation (tangential), we used a color scheme that allows us
to visualize both with the same color. In the plot shown in
Fig.~\ref{polvec}, yellow and green colors correspond to radial magnetic field
orientations, while blue colors indicate tangential ones.  It is clear
that the overwhelmingly radial direction in the bright lobes is not
preserved in the SE and NW, where the orientation is predominantly
tangential.  Even in the lobes, there are some departures from the
radial direction, mainly at the rim.

Furthermore, we selected only those pixels with radial field orientations within
$\pm 14^\circ$, which is the error in the position angle (Section 2), and 
plotted them in green in Fig. \ref{overlap}.  We also selected vectors parallel
to the Galactic Plane, which lies at a fixed angle of $60^\circ$ (counterclockwise from North) 
as projected in this image, and overplotted them in red. In both cases, the scale is such that pixels 
in the limits of the tolerance interval are displayed with the lowest intensity.
Finally, pixels not belonging to any of these two groups are overplotted in 
blue. In this case, the faintest pixels correspond to angles closest to the 
Galactic Plane direction. Along this direction (yellow line in Fig. 
\ref{overlap}), pixels appear in yellow, clearly indicating that the magnetic 
vectors there fulfill both criteria. 

\begin{figure}[ht]
  \centering
  \subfigure[]{
    \label{swcomp}
    \includegraphics[width=6.8 cm]{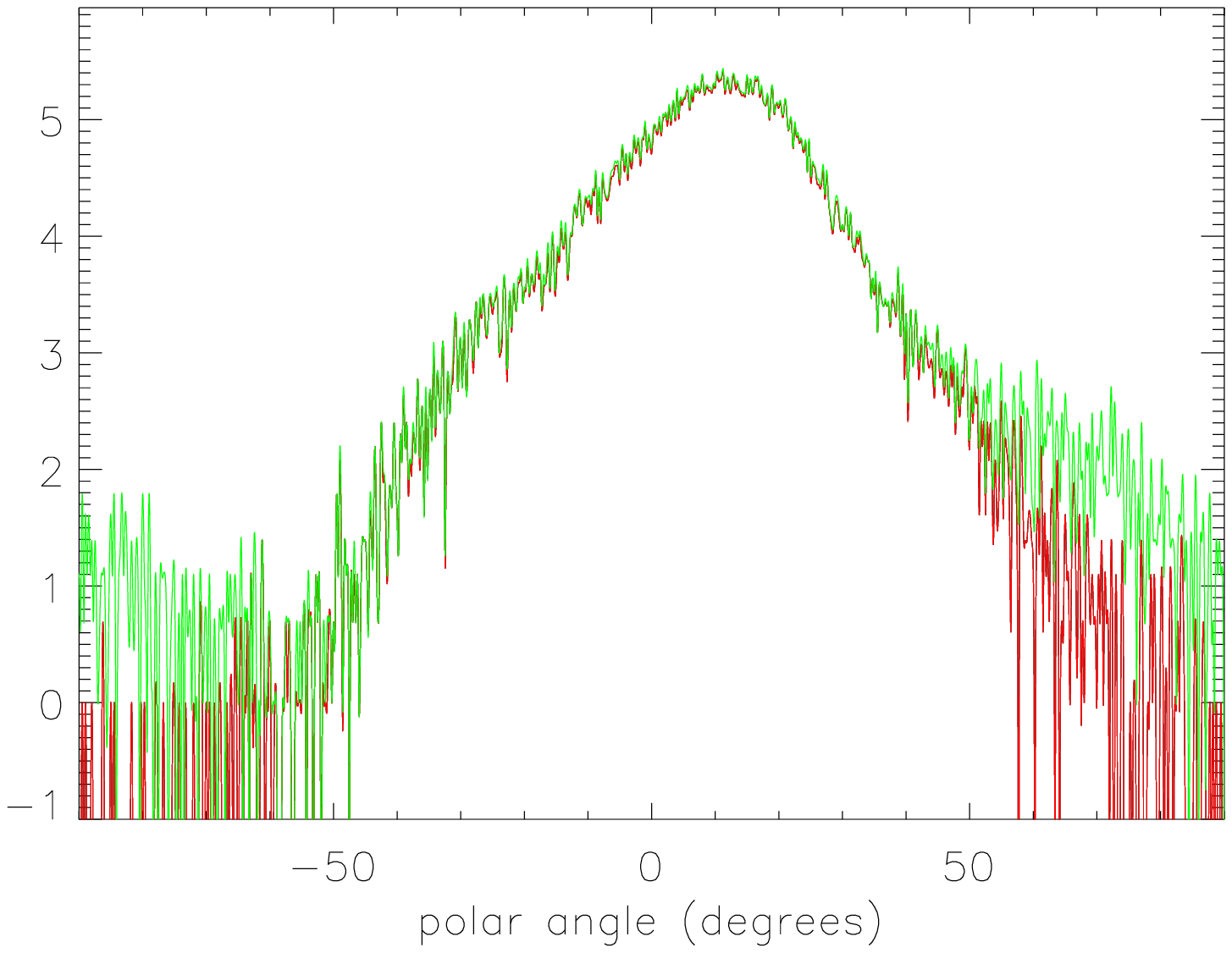}
  }
  \subfigure[]{
    \label{swtan}
    \includegraphics[width=7 cm]{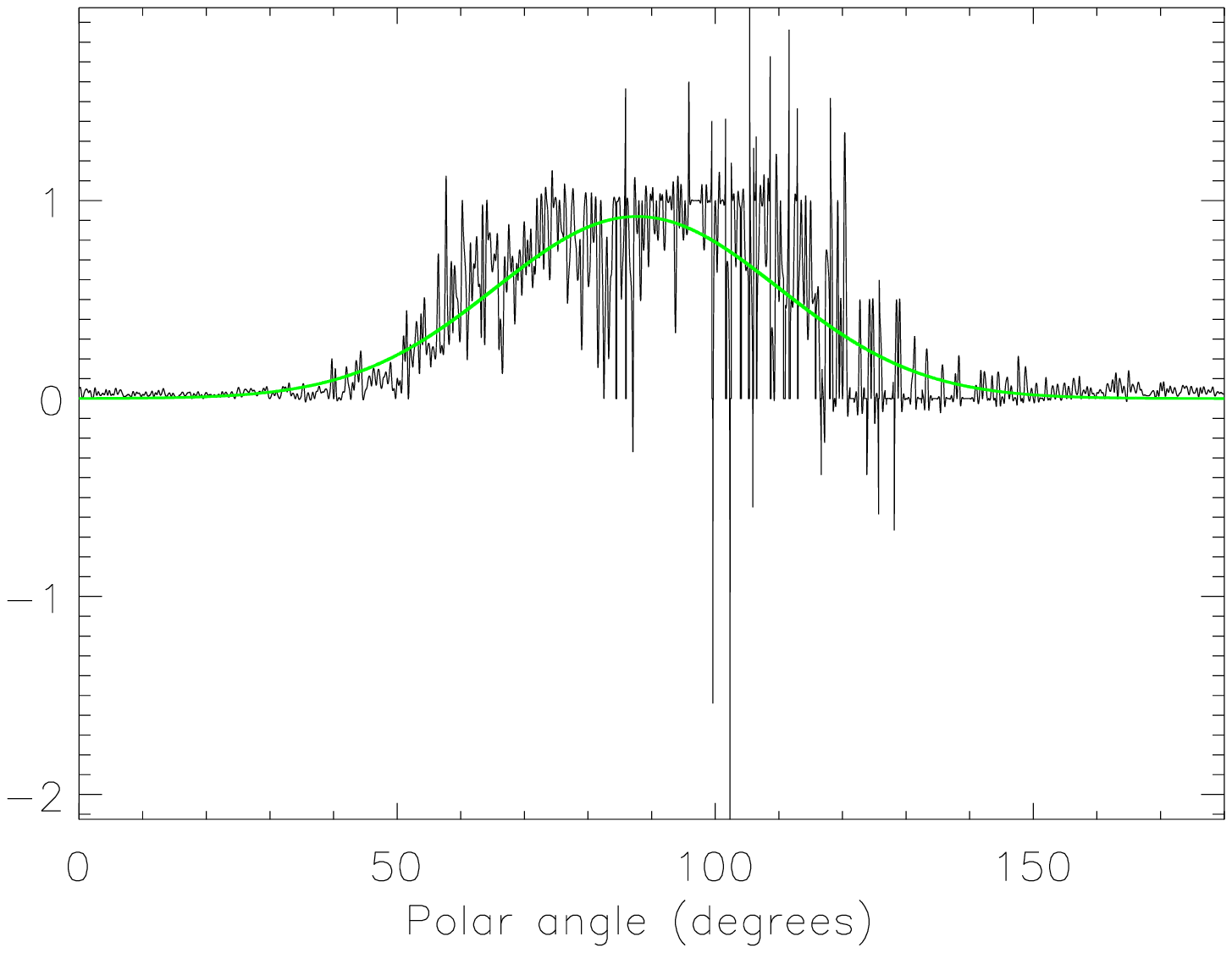}
  }
  \caption{(a) Logarithmic distribution of the direction of magnetic vectors in 
the SW sector based on a polar coordinate reference system, extracted from 
pixels within a radial wedge defined between 170\degb and 250\degb in azimuth, 
and from 540 to 860 arcsec in radius (red) or to 920 arcsec in radius (green). 
(b) Difference polar-referenced angle distribution (see text). }
  \label{sw}
\end{figure}

\begin{figure}[ht]
  \centering
\resizebox{\hsize}{!}{\includegraphics{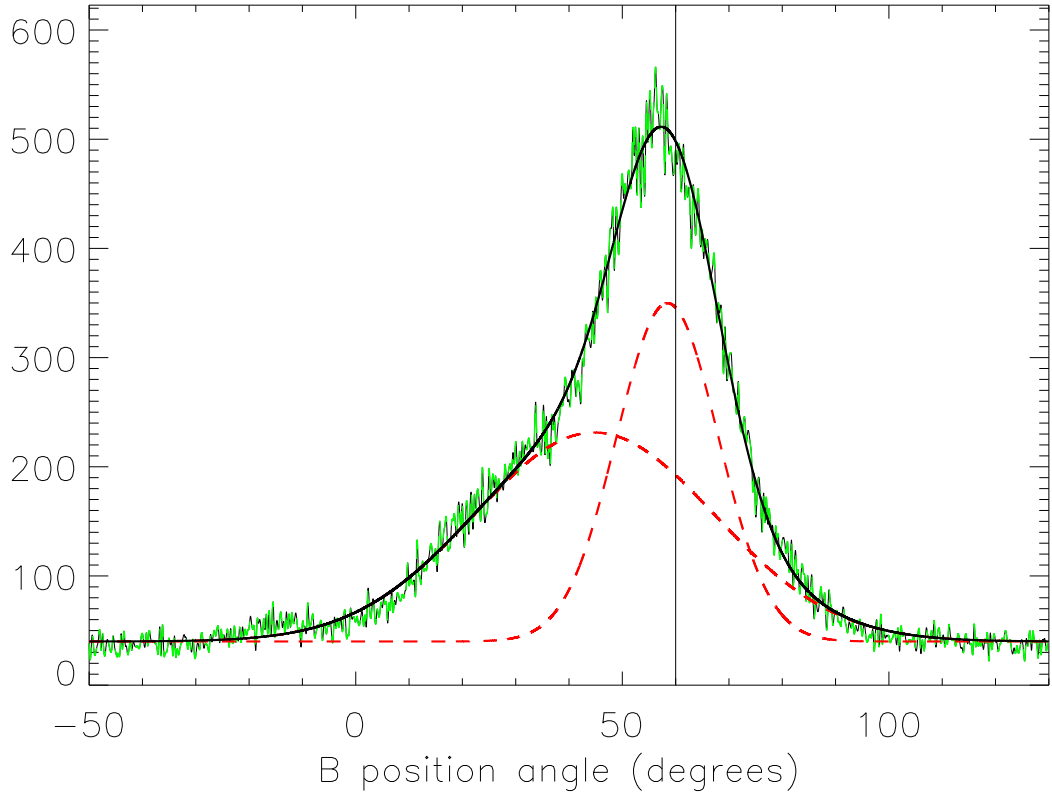}}
  \caption{Distribution of the position angle of the magnetic field measured 
with respect to a frame fixed to the celestial coordinates (green curve). The 
best fit is represented by a black solid line, which is the result of the 
addition of two Gaussian distributions (red, dashed lines) plus a constant. The 
vertical solid line corresponds to the Galactic Plane tilt in the direction of 
SN 1006.} 
\label{galpl}
\end{figure}

\begin{figure}[ht]
  \centering
\resizebox{\hsize}{!}{\includegraphics{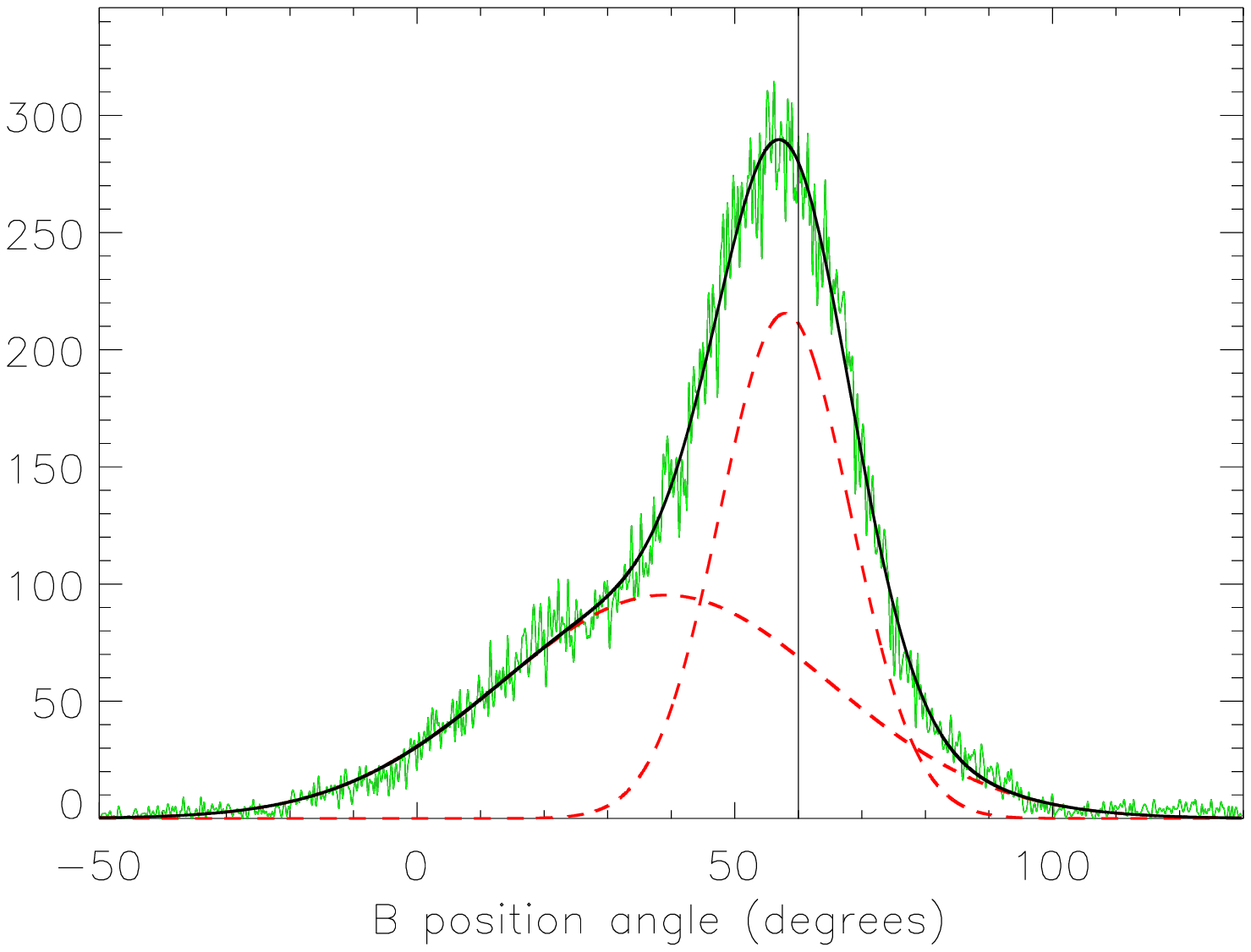}}
  \caption{Same as Fig.~\ref{galpl} but considering only the two bright lobes.}
\label{galpl2}
\end{figure}

We examined the distribution of polar-referenced magnetic field direction (Fig. \ref{polvec}) solely using information 
from pixels within the NE, SW and SE regions of SN 1006.  We extracted the information from the bright lobes using pixels 
within two radial wedges with azimuth ranges from 20\degb to 100\degb (NE) and 170\degb to 250\degb (SW),  and radial ranges 
from 540 to 860 arcsecs from the center.  We found that the distributions peak at 9\degb in the NE lobe and -4\degb in the 
SW lobe, with standard deviations, $\sigma$, of 20\degb and 25\deg, respectively.  We extracted information from pixels 
in the SE using a radial wedge with an azimuth range from 113\degb to 147\degb, and a radial range from 865 to 920 arcsec 
from center.  We shifted the difference angle range from [$-90$\deg, 90\deg] to [0\deg, 180\deg], a more appropriate range 
for identifying magnetic fields that are tangential (purely tangent would then be 90\deg).  The distribution in the SE 
peaks at $\sim 112$\deg, with a standard deviation of 18\deg, clearly consistent with a tangential orientation.

A surprising result is obtained if the outer radius of the SW lobe
sector is extended so as to include the outer rim (see Fig. \ref{sw}). 
In Fig.~\ref{swcomp}, the logarithmic distribution of polar-referenced field angles, 
including only the inner sector (radius of 860 arcsec) is plotted in 
red, while the green line displays the distribution at a radius of 920 arcsecs. 
While the distribution is the same for polar referenced angles between approximately $-50$\degb and 50\deg, there is a clear excess of more oblique vectors when the region with radii between 860 and 920 arcsecs 
is included. To further highlight this trend, in Fig.~\ref{swtan} we plot 
the polar referenced angle distribution from only the radial range between 860 and 920 
arcsec (with the distribution shifted to the range [0\deg , 
180\deg ] as before). A simple Gaussian fit (albeit crude given the evident 
non-Gaussian nature of the distribution) gives a peak at $\sim 88$\degb with a 
deviation of 30\deg.  Although more sensitive polarization observations are
desirable to confirm this result, it is worth noting that indications
of a tangential magnetic field at the outer edge have also been
reported in the Cassiopeia A SNR \citep{got+01}.

Finally, we examined the distribution of magnetic field orientation
with respect to the direction of the Galactic Plane. The distribution,
shown in Fig.~\ref{galpl}, can be reasonably well fitted by the sum of
a constant term and two Gaussians (red, dashed lines): a broad one
centered at $45^\circ$ and a narrow one centered at $58^\circ$. The
result is represented by the green line.  The vertical solid line
overlaid at $60^\circ$ corresponds to the position angle of the
Galactic Plane in this direction. In Fig.~\ref{galpl2}, we consider
only the two bright lobes, which provide a similar result: 47\% of the
magnetic vectors are clustered around $58^\circ$, while the rest fall
in a broad component centered at $\sim 40^\circ$.

\section{Discussion}

\citet{milne87} reviewed observations of polarized radio emission for
27 SNRs and by examining their projected magnetic fields, concluded
that young remnants have radial magnetic fields while well-defined
tangential fields are seen only in older SNRs. Subsequent observations
\citep[e.g.,][]{tycho97,land+99,kes69} have supported this picture. In
the case of SN 1006, the magnetic field orientation could only be
measured on the bright lobes \citep[][and references therein]{rg93}
and, as expected for a young SNR, was found to be radial. These
studies left open the question of whether fainter regions of the shell
had a radial field as well.

If the projected magnetic field were radial all around the shell, then
the measured position angles with respect to North should be uniformly
distributed between $-$90\degb and 90\deg.  However, considering that
most of the information on the magnetic field in SN 1006 comes approximately 
from the first and third quadrants (the NE and SW bright lobes), there 
will be a bias in the distribution favoring position angles between 
0\degb and 90\deg. While the broad Gaussian centered at 45\degb in 
Fig.~\ref{galpl} is not conclusive support for this model, the narrow 
component near $60^\circ$ is clearly at odds with it.  When we reproduce the 
histogram for azimuthal sectors of $10^\circ$ width over the lobes, 
the fixed angle component near $60^\circ$ is found in most, but not
all, of the sectors.

Examination of Fig. \ref{overlap} clarifies the situation, showing
clearly that the pattern of magnetic field vector orientations is not
uniform. Patches with a radial orientation (colored green) are found
all along the northern lobe and partially in the southern lobe.  On
the other hand, regions with vectors parallel to the Galactic Plane
(colored red) are found over practically all of the shell. (The pixels
colored yellow are located where the local radial direction happens to
match the fixed field direction.)  The patches with radial and fixed
field directions are interspersed throughout the bright lobes. There
is no correlation between the magnetic field orientation and the
fractional polarization in these patches. At the
outermost edge of the remnant's shell, several azimuthal sectors display
orientations (coded in blue) which are neither radial nor parallel to
the Galactic Plane, possibly pointing to the addition of a radial and
a fixed Galactic component.

We have established that an important fraction of the magnetic field vectors 
in SN 1006 are parallel to a fixed angle near $60^\circ$. This
is particularly clear in the SE sector, where the magnetic field 
vector distribution is strongly dominated by a fixed field component at an 
angle of $\sim$ $54^\circ$ and a broadening width of $\sigma \approx 
10^\circ$. We draw the conclusion that the fixed angle component indicates 
the direction of the ambient magnetic field projected onto the plane of the 
sky, which in this case runs nearly parallel to the Galactic Plane. This
allows us to answer the long-lasting question concerning which of the
two competing models explains the bright lobes, $i.e.$, whether they
arise from a limb-brightened equatorial belt or two polar caps.  Our
study finds that the two bright lobes in SN 1006 are actually polar
caps.

The linear polarization fraction of an optically thin source emitting
synchrotron radiation in a uniform field is limited by the following
relation: $p \le \left( { \alpha + 1 })/( {\alpha + 5/3 }
\right)$. The mean spectral index for SN 1006 is $\alpha$ = 0.6, so
that the expected maximum value for $p$ is 70.5\%. Such high values
are not observed for SNRs. In general, $p$ is measured to be 10\% --
15\%, with higher values (35\% -- 50\%) in just a few exceptional
cases \citep{land+99,dms00,rg07}. The measured fractional polarization
of the SW and NE lobes is about 17\%, in agreement with previous
polarization studies towards SN 1006 but far from the theoretical
maximum.  According to \citet{burn66}, this means that only 24\% of
the magnetic energy is in an ordered component. Therefore, even though
the magnetic fields appear ordered at the level of our image
resolution, the low fractional polarization indicates more generally
that the fields on smaller scales are highly disorganized in these
regions.

In contrast, the SE region of faint radio and non-existent X-ray synchrotron 
emission shows a high fractional polarization compatible with the theoretical
expected value. This requires the emission region to host a highly ordered
magnetic field, likely just the swept-up ambient field \citep[see][]{cassam+08}.
Figs.~\ref{magvec} and ~\ref{polvec} show that the magnetic field is mostly 
tangential at this faint region. This orientation was derived assuming a 
uniform RM for the whole SNR. Were the RM half the average value in this 
region, the vectors would rotate counter-clockwise by $\sim 10^\circ$, but 
would still be close to the external magnetic field direction estimated from 
the bright lobes.

We can therefore link efficient particle acceleration and disordered, amplified
magnetic fields with quasi-parallel shocks in the bright SW and NE lobes of
SN 1006. Likewise, we link inefficient particle acceleration and highly-ordered
swept-up magnetic fields with quasi-perpendicular shocks in the faint SE rim.
This scenario is also in agreement with the recent detection of very high energy
$\gamma$-rays at the lobes \citep{hess10}. \citet{pet+12} consider the 
coincidence between the radio, X-rays and $\gamma$-ray emission as possible
evidence for a leptonic origin of the $\gamma$-rays, and set upper limits to 
the magnetic field strength of a few hundred micro-gauss, which are less 
constraining than previous determinations \citep[see][and references 
therein]{pet+12}.

The radial magnetic fields usually observed in young SNRs have been ascribed to 
amplification along the Rayleigh-Taylor fingers developed at the interface 
between the shocked ambient medium and the ejecta \citep[e.g.,][]{junorman}.  
However, we have found that the radial distribution is only partially observed 
in the bright lobes of SN 1006.  \citet{dpt04} solved the nonlinear problem of 
the generation of a magnetic field by particles accelerated in a MHD shock 
front. Their model is based on the assumption of uniform stationary particle
injection, which is probably suitable for SN 1006 due to the low density and 
homogeneity of the ambient medium, far as it is from the Galactic Plane. They 
found that when the ambient magnetic field has a component perpendicular to the 
shock front, the parallel component is destroyed by the secondary field 
generated by the accelerated particles, resulting in a radial field 
distribution. Our results for the bright lobes suggest that this rearrangement 
is not complete and a fraction of the original field survives as inferred from
the peak near $60^\circ$ in the distribution of magnetic field directions. On
the other hand, if the shock front were parallel to the ambient magnetic field, 
\citet{dpt04} find a second possible solution in which the field is not 
suppressed but, instead, is amplified and can be two orders of magnitude 
stronger than the mean value measured in the cold and warm phases of the ISM. 
We offer the following comments, based on the properties of the SE sector of 
SN 1006, where the observed field is nearly parallel to the shock front. First, 
\citet{cassam+08} showed that the faint radio surface brightness there could be 
accounted for by a small injection rate of particles into the acceleration 
process (corresponding to $<$ 10\% of the shock energy going into accelerated
particles) and shock compression by a factor of 4 of a typical 3 $\mu$G ambient 
ISM magnetic field. In other words no magnetic field amplification appears to 
be required in the SE sector. Second, the high degree of radio polarization we 
find in the current study requires that the post-shock magnetic field be highly 
uniform. The interpretation we favor is that particle acceleration and the
generation of turbulent magnetic field are inefficient at the SE rim of SN 
1006. 

We take the fixed angle component observed almost all along the SNR shell as a
strong indication of the ambient magnetic field direction.  The
prevalence of this direction at the highly polarized SE arc, which can
be explained by the second solution of \citet{dpt04}, further supports
this hypothesis. The patchy distribution of regions with either fixed
angle or radial magnetic vectors, which is a surprising result of our 
study, remains to be understood.

\section{Conclusions}

We have presented a new radio polarization study of SN 1006 that has
unveiled new features related to the polarization fraction and
magnetic field direction of both the SNR and the surrounding
ISM. These findings have implications for DSA theory. We found
that the brightest radio, X-ray, and TeV $\gamma$-ray features have
the lowest polarization fractions, indicating the presence of a
disordered, turbulent magnetic field precisely in the regions where
particle acceleration is most efficient. Moreover, the SE region of
the shell, with very faint synchrotron emission, shows a fractional
polarization close to the theoretical maximum value for an optically thin
source, indicating a highly ordered magnetic field at a location where
particle acceleration is very inefficient \citep{cassam+08}.  While most of
the SNR shell appears to have a radial magnetic field distribution, we have 
found that a significant portion of the distribution share a single angle 
of $\sim 60^\circ$, which in turn coincides with the Galactic Plane direction. 
In the SE sector the distribution of magnetic field directions is consistent 
with a single component at an angle of $\sim 55^\circ$.  Overall, the 
evidence points to an ambient magnetic field roughly parallel to the Galactic
Plane, of which the SNR retains some knowledge even after the passage
of the shock front. The portions of SN 1006 where the shocks are
quasi-parallel are more efficient at producing particle acceleration
and generating magnetic turbulence; elsewhere where the shocks are
quasi-perpendicular particle acceleration is inefficient and there is
little turbulent magnetic field generation. Finally, given these
results, considerations of geometrical symmetry clearly establish
that, rather than being the limb-brightened edges of an equatorial
belt, the bright synchrotron emitting lobes in SN 1006 are polar caps.

\section*{Acknowledgements}

The authors acknowledge the anonymous referee for comments that improved the 
presentation of our results.  We also thank Roger Blandford and Stephen Reynolds for their
critical reading of the manuscript.  E.~M.~R.\ is partially supported by grants PIP 114-200801-00428
(CONICET) and UBACyT 20020090200039. D.~A.~M.\ acknowledges support
from a Furman University RPG Grant.  The National Radio Astronomy Observatory is a facility of the National Science Foundation operated under cooperative agreement by Associated Universities, Inc.
The Australia Telescope is funded by the Commonwealth of Australia for
operation as a National Facility managed by CSIRO.  This research has
made use of the NASA/IPAC Extragalactic Database (NED) which is
operated by the Jet Propulsion Laboratory, California Institute of
Technology, under contract with the National Aeronautics and Space
Administration.

\bibliography{reyn0824_rev3}

\clearpage

\end{document}